\documentclass[final,5p,times,twocolumn,authoryear]{elsarticle}

\usepackage{geometry}
\usepackage{amssymb,amsmath,verbatim,mathtools,needspace,enumitem,etoolbox,graphicx,physics,microtype,afterpage,bigints,gensymb,tabularx,xspace}

\usepackage[dvipsnames, usenames]{xcolor}
\usepackage{booktabs}
\usepackage{array}
\definecolor{linkcolor}{rgb}{0.0,0.3,0.5}
\definecolor{dodgerblue}{HTML}{1E90FF}
\usepackage[unicode, colorlinks=true, linkcolor=linkcolor, citecolor=linkcolor, filecolor=linkcolor,urlcolor=linkcolor]{hyperref}
\hypersetup{
    pdftitle={Latin hypercube sampling to reduce the variance in Monte Carlo integration: a practical guide and an application to gravitational-wave astronomy},
    pdfauthor={Matteo Boschini and Davide Gerosa}}
\usepackage[all]{hypcap}
\usepackage[T1]{fontenc}
\usepackage[utf8]{inputenc}
\usepackage{orcidlink}

\definecolor{olhsSymA}{RGB}{214,234,248}   %
\definecolor{olhsSymB}{RGB}{252,228,214}   %
\definecolor{olhsSymC}{RGB}{226,240,217}   %

\newlength{\olhsTileWidth}
\newcommand{\olhsTile}[2]{%
  {\setlength{\fboxsep}{1.5pt}\colorbox{#1}{\makebox[\olhsTileWidth][c]{\strut#2}}}}

\newcommand{\olhsA}[1]{\olhsTile{olhsSymA}{#1}}
\newcommand{\olhsB}[1]{\olhsTile{olhsSymB}{#1}}
\newcommand{\olhsC}[1]{\olhsTile{olhsSymC}{#1}}

\begin{document}

\begin{frontmatter}

\title{Latin hypercube sampling to reduce the variance in Monte Carlo integration:
\\
a practical guide and an application to gravitational-wave astronomy}

\author[milan,infn]{Matteo Boschini$\,$\orcidlink{0009-0002-5682-1871}}
\ead{m.boschini1@campus.unimib.it}
\author[milan,infn]{Davide Gerosa$\,$\orcidlink{0000-0002-0933-3579}}

\affiliation[milan]{organization={Dipartimento di Fisica ``G. Occhialini'', Universit\`a degli Studi di Milano-Bicocca},
    addressline={Piazza della Scienza 3},
    city={Milano},
    postcode={20126},
    country={Italy}}
\affiliation[infn]{organization={INFN, Sezione di Milano-Bicocca},
    addressline={Piazza della Scienza 3},
    city={Milano},
    postcode={20126},
    country={Italy}}

\begin{abstract}

In applied statistics, likelihoods involving high-dimensional integrals are often approximated via Monte Carlo summation. Controlling the resulting numerical variance is crucial for reliable inference. We provide a practical introduction to Latin hypercube sampling (LHS) and its extension based on orthogonal arrays (OA-LHS) as strategies to reduce this variance by stratifying the sampling distribution. LHS asymptotically removes variance contributions from individual coordinates, while strength-2 OA-LHS also removes those from pairwise interactions,  yielding orders-of-magnitude improvements over plain Monte Carlo estimators for suitable integrands.
 As a case study, we investigate their application to the selection function entering hierarchical inference of gravitational-wave populations. At approximately one million injections, we find only a marginal reduction in variance, with low statistical significance.
Nevertheless, the computational cost of constructing these designs is negligible relative to evaluating gravitational-wave injections, making even marginal gains worth considering for future applications. This practical introduction to LHS and OA-LHS  offers a starting point for other applications in physics and astronomy involving computationally expensive Monte Carlo calculations.

\end{abstract}

\end{frontmatter}

\section{Introduction}

Monte Carlo summation is a powerful technique for approximating high-dimensional integrals, often scaling more favorably than conventional grid-based quadrature \citep{caflisch1998monte}. Such integrals arise naturally in hierarchical Bayesian inference, where likelihoods involve marginalization over parameters at lower levels of the hierarchy. Evaluating these integrals with Monte Carlo sums introduces numerical noise into the likelihood estimate, which can hinder sampler convergence or bias inference. Estimating and controlling this uncertainty is therefore crucial to prevent numerical fluctuations from obscuring the posterior structure.

Whenever computationally feasible, the variance of Monte Carlo estimators can be reduced simply by increasing the number of samples. Less trivially, the sampling strategy can be optimized to reduce the variance at a fixed sample size. This is the domain of simulation design \citep{garud2017design,pronzato2012design}, which is concerned with optimizing the placement of sample points within a given parameter space. In this context, Latin hypercube sampling (LHS) is a popular space-filling strategy for drawing randomized multidimensional samples with uniform stratification in each one-dimensional marginal~\citep{eglajs1977new,ef76b040-2f28-37ba-b0c4-02ed99573416,HELTON200323}.
Generalizations such as orthogonal-array Latin hypercube sampling (OA-LHS) extend this stratification to higher-dimensional marginals~\citep{c7ab10ee-a62f-3f85-b103-7b6e0971f8b6, Tang01121993}. Crucially, under appropriate regularity conditions, the asymptotic variance of an LHS-based Monte Carlo estimator is \emph{mathematically guaranteed} to be no greater than that of a standard Monte Carlo estimator~\citep{fbe9475b-1452-3229-a386-04540442a9c3,61820f3e-e4da-3e76-b859-f9f34a370a7d}.

This property makes LHS approaches particularly useful in applied statistics, where reducing the variance of integral estimates entering the likelihood can be crucial. In this paper, we provide an introduction ``for physicists'' to Monte Carlo integration with LHS and OA-LHS, together with an application to our field of research, gravitational-wave (GW) astronomy.

In GW astronomy, hierarchical Bayesian statistics is used to estimate the population properties of merging black holes and neutron stars. The resulting population likelihood combines single-event likelihoods with a population model and includes a normalization term that accounts for selection effects~\citep{2019MNRAS.486.1086M,2022hgwa.bookE..45V}. Evaluating this likelihood requires the computation of $N_{\rm obs}+1$ integrals, where $N_{\rm obs}$ is the number of events in the GW catalog. Of these, $N_{\rm obs}$ integrals arise from marginalizing over the parameters of each individual source, while the additional integral captures the detection efficiency and thus encodes our model of the detectors. This latter integral enters the denominator of the likelihood raised to the $N_{\rm obs}$-th power, amplifying the impact of numerical uncertainty.

In GW population fits, approximating integrals with Monte Carlo sums is particularly convenient because it allows existing data products to be reused efficiently: single-event posterior samples obtained with convenient but unphysical priors and sensitivity injections. A substantial body of literature has been devoted to quantifying the variance these approximations introduce into the GW population likelihood~\citep{2019RNAAS...3...66F,2022arXiv220400461E,2023MNRAS.526.3495T,2025PhRvD.111j3012M,2025arXiv250907221H,2026PhRvD.113j3025W}. %
In practice, obtaining stable results in current, state-of-the-art analyses requires excising regions of parameter space where the Monte Carlo variance is deemed too large. While this is often referred to, at least colloquially, as a ``data-dependent prior,'' it in fact corresponds to modifying the likelihood, invalidating the assumptions used to derive the likelihood itself. %
At present, this is the only workable solution. %

Of the $N_{\rm obs}+1$ integrals entering the population likelihood, the one that dominates the overall variance is the selection function~\citep{2023MNRAS.526.3495T}. This quantity encodes the fraction of detectable objects in a population with given parameters. Let $\theta$ denote the single-event parameters (e.g.\ the masses and spins of black holes) and $\lambda$ the population parameters sampled in the hierarchical analysis. The selection function is then given by
\begin{equation}
\label{xiintegral}
\xi(\lambda) = \int p_{\rm pop}(\theta | \lambda)\, p({\rm det} | \theta)\, \dd \theta \, ,
\end{equation}
where $p_{\rm pop}(\theta | \lambda)$ is the assumed population model and $p({\rm det} | \theta)$ is the probability of detecting a source with parameters $\theta$ during a given observing campaign.
This integral is usually computed via importance sampling from a predefined set of pipeline injections---simulated signals added to detector data and searched for with the same pipelines used for real events \citep{2018CQGra..35n5009T}. The resulting Monte Carlo estimator is
\begin{equation}
\label{xihat}
\xi(\lambda)
\sim
\left\langle
\frac{p_{\rm pop}(\theta_i | \lambda)}{p_{\rm inj}(\theta_i)}
\, p({\rm det} | \theta_i)
\right\rangle_{\theta_i \sim p_{\rm inj}} \, ,
\end{equation}
where $p_{\rm inj}(\theta)$ is the distribution used to generate the injections. Generating these injections requires significant computational and bookkeeping effort; in current analyses~\citep{2026arXiv260527226T}, selection effects are estimated using $\sim4\times10^{\rm 8}$ %
such samples, drawn from a distribution $p_{\rm inj}$ that was carefully constructed to optimize downstream applications~\citep{2025PhRvD.112j2001E}.

Several approaches have been proposed to make the treatment of selection effects more efficient. These include machine-learning emulators~\citep{2020PhRvD.102j3020G,2022ApJ...927...76T,2024PhRvD.110l3041C,2026arXiv261000306G}, which aim to learn a smooth approximation to the detection probability $p({\rm det} | \theta)$, as well as semi-analytic approaches~\citep{1993PhRvD..47.2198F,2023PhRvD.108d3011E,2024CQGra..41l5002G,2025CQGra..42d5008L}, which exploit simplified models of detector sensitivity to compute the selection function more efficiently. Alternatively, one can target the observed distribution instead of the intrinsic one~\citep{2026PhRvD.113h3006T}. While promising, these methods introduce additional modeling assumptions and complexities, and their accuracy must be carefully validated.

The sampling-design perspective introduced above offers a complementary route: retain the Monte Carlo estimator and improve how injections are drawn from $p_{\rm inj}$. LHS and OA-LHS impose stratification on these draws, which can reduce fluctuations associated with uneven coverage of the sampling space at negligible additional cost compared to generating and analyzing the injections. The resulting variance reduction depends on the structure of the integrand, making the GW selection function a useful test of both the benefits and the limitations of these methods.

We first review the theory behind LHS and OA-LHS (Sec.~\ref{mcintegrals}) and illustrate these methods using toy models (Sec.~\ref{examples}). We then quantify their effect on the variance of the GW selection-function estimate (Sec.~\ref{selectionfunction}) and summarize our results (Sec.~\ref{conclude}).

\section{Monte-Carlo integration with Latin hypercubes}
\label{mcintegrals}

\begin{figure*}[t]
	\centering
	\includegraphics[width=0.8\columnwidth]{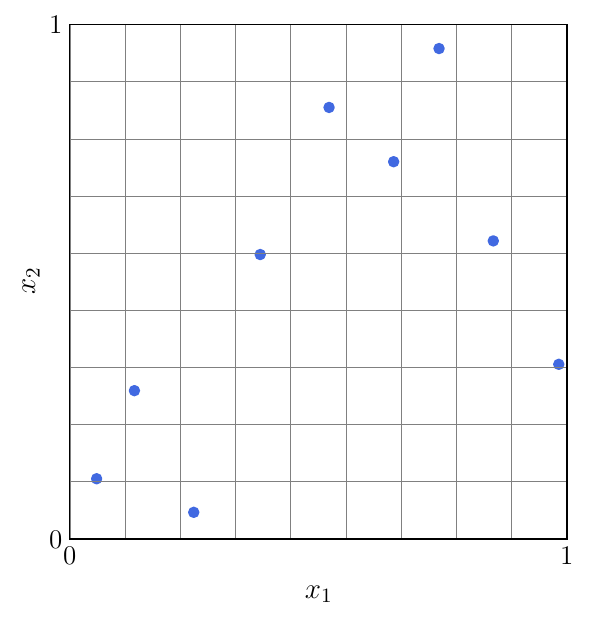}
	$\qquad$
	\includegraphics[width=0.8\columnwidth]{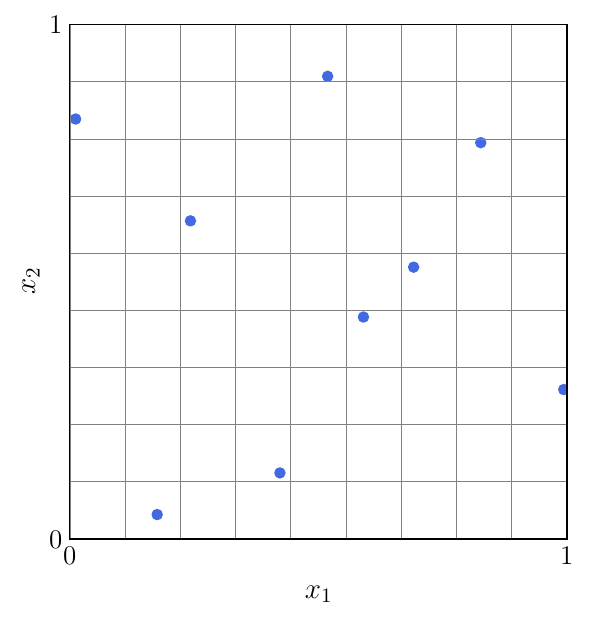}
	\caption{The left panel shows an example of a Latin hypercube set LHS$\left(9, 2\right)$. The right panel shows instead an orthogonal array Latin hypercube set OA-LHS$\left(9, 2\right)$.}
	\label{fig:one}
\end{figure*}

\subsection{Latin hypercube sampling}
LHS is a technique that allows drawing $N$ points in the unit hypercube in $d$ dimensions $[0,1)^d$ while guaranteeing uniform distributions in each one-dimensional projection.

A practical construction of an LHS design with $N$ points in the unit hypercube $[0,1]^d$ is as follows:
\begin{enumerate}
\item Divide each dimension into $N$ disjoint bins $[i/N, (i+1)/N)$, with $i=0,\dots,N-1$.

\item Select one bin uniformly at random in each dimension, independently across dimensions, and draw a point uniformly within the resulting $d$-dimensional cell.

\item For each subsequent point, repeat the previous step, restricting the selection in each dimension to bins that have not been selected previously.

\item Continue until $N$ points have been drawn.

\end{enumerate}

This ensures that, in the resulting set LHS$\left(N, d\right)$, each bin in each dimension contains exactly one sample. The left panel of Fig.~\ref{fig:one} shows a particular realization of LHS$\left(9, 2\right)$.

For a more efficient algorithm, such as that implemented in SciPy \citep{2020NaMet..17..261V}, the same construction can be reformulated as follows:

\begin{enumerate}
\item
Divide each coordinate axis into $N$ intervals of equal width $1/N$, as above.

\item Draw one value uniformly at random from each interval, independently for every interval and coordinate. This produces $d$ lists, each containing $N$ values.

\item Randomly permute each list, choosing the permutations uniformly and independently across coordinates.

\item Assemble the sample points by taking the $i$-th entry from each of the $d$ shuffled lists as the coordinates of the $i$-th point, for $i=0,\ldots,N-1$.
\end{enumerate}
In symbols, the LHS elements are given by
\begin{equation}
\label{plainLHS}
	u_{ij}=\frac{\pi_j(i)+V_{ij}}{N},
	\qquad
	V_{ij}\sim\mathcal{U}[0,1),
\end{equation}
where, for each dimension $j$, $\pi_j$ is a random permutation of ${0,\dots,N-1}$.

Multiple LHS sets can be constructed for a given number of samples $N$ and hypercube dimensionality $d$. Individual designs can still exhibit poor multidimensional coverage, including strong spurious correlations between dimensions or clustering of points, leaving room for further optimization. %
Rank-correlation control can be used to reduce pairwise correlations or impose a specified degree of correlation~\citep{Iman01011982, Owen01121994, MAINIK2015197}, and distance-based or discrepancy-based metrics can serve as optimality criteria for selecting or optimizing a given set. The maximin distance~\citep{JOHNSON1990131}, $\phi_p$~\citep{MORRIS1995381}, and centered discrepancy~\citep{ZHOU2013283} are examples of possible metrics. Stochastic optimization algorithms, such as the enhanced stochastic evolutionary algorithm~\citep{JIN2005268}, can be used to search for a particular Latin hypercube realization optimizing these criteria. However, these optimizations become increasingly expensive for large $N$, since most metrics are distance-based and scale as~$\mathcal{O}\!\left(N^2\right)$.

\subsection{Orthogonal arrays}

A more promising approach to obtain an optimized LHS set is based on orthogonal arrays~\citep{c7ab10ee-a62f-3f85-b103-7b6e0971f8b6, Tang01121993}. By definition, an orthogonal array of type $\left(N,d,q,t\right)$ is an $N\times d$ matrix whose entries are chosen from the set $\{0,\dots,q-1\}$ such that in every $N\times t$ sub-matrix each of the $q^t$ possible $t$-tuples appears equally often, namely $\lambda=N/q^t$ times~\citep{0a7eac65-4c60-3844-ab27-7dd6b1426168,hedayat1999orthogonal}. The parameter $t$ is called the strength of the orthogonal array.
The development of algorithmic strategies to build arrays with high strengths is a topic of active research. In this work, we focus on the construction of a Latin hypercube starting from an orthogonal array with strength $t=2$. %

\citet{4d08eb2d-e239-34a6-b395-526fdad615ca} and \citet{24bb4866-58f9-3a8a-9dbf-7dc49ed87eae} introduced a procedure to obtain OA$\left(p^2,p+1,p,2\right)$, where $p$ is a prime number. From this orthogonal array, one then builds an OA-LHS set with $N=p^2$ samples in $d\leq p+1$ dimensions.%

The construction proceeds as follows:

\begin{enumerate}

\item As a starting point, consider the first two columns of a table, $a$ and $b$, and build  $p^2$ rows with pairs $\left(a_i,b_i\right) \in \mathbb{Z}^2_p$, with $\mathbb{Z}_p=\{0,\dots,p-1\}$. Then the remaining $p-1$ columns are built as $a+cb \pmod{p}$ with $c \in\{1,\dots,p-1\}$. %
For instance, the following table showcases the base OA$\left(9,4,3,2\right)$ corresponding to the case $p=3$.  %
\begin{center}
        \begin{tabular}{ccccc}
                $i$ & $a$ & $b$ & $a+b$ & $a+2b$ \\
                \midrule
                0 & \olhsA{0} & \olhsA{0} & \olhsA{0} & \olhsA{0} \\
                1 & \olhsA{0} & \olhsB{1} & \olhsB{1} & \olhsC{2} \\
                2 & \olhsA{0} & \olhsC{2} & \olhsC{2} & \olhsB{1} \\
                3 & \olhsB{1} & \olhsA{0} & \olhsB{1} & \olhsB{1} \\
                4 & \olhsB{1} & \olhsB{1} & \olhsC{2} & \olhsA{0} \\
                5 & \olhsB{1} & \olhsC{2} & \olhsA{0} & \olhsC{2} \\
                6 & \olhsC{2} & \olhsA{0} & \olhsC{2} & \olhsC{2} \\
                7 & \olhsC{2} & \olhsB{1} & \olhsA{0} & \olhsB{1} \\
                8 & \olhsC{2} & \olhsC{2} & \olhsB{1} & \olhsA{0} \\
        \end{tabular}
\end{center}
Hereafter, we refer to symbols entering the array as $k_i$. Their numerical value will change as the construction proceeds. We track entries in by coloring the cells in the tables below according to this first step.

	\item Following \citet{c7ab10ee-a62f-3f85-b103-7b6e0971f8b6}, each column $j$ is relabeled by an independent random permutation $\pi_j$ of the symbol set $\{0,\dots,p-1\}$: every occurrence of symbol $k$ in column $j$ is replaced by $\pi_j(k)$, i.e.\ $\tilde{A}_{ij} = \pi_j(A_{ij})$. The resulting scrambled array is shown below. For instance, in this example, column 0 has the following map: $0 \mapsto 2$, $1 \mapsto 0$, $2 \mapsto 1$. %
\begin{center}
	                \begin{tabular}{ccccc}
                        $i$ & $\tilde A_{i0}$ & $\tilde A_{i1}$ & $\tilde A_{i2}$ & $\tilde A_{i3}$ \\
                        \midrule
                        0 & \olhsA{2} & \olhsA{1} & \olhsA{0} & \olhsA{2} \\
                        1 & \olhsA{2} & \olhsB{2} & \olhsB{2} & \olhsC{0} \\
                        2 & \olhsA{2} & \olhsC{0} & \olhsC{1} & \olhsB{1} \\
                        3 & \olhsB{0} & \olhsA{1} & \olhsB{2} & \olhsB{1} \\
                        4 & \olhsB{0} & \olhsB{2} & \olhsC{1} & \olhsA{2} \\
                        5 & \olhsB{0} & \olhsC{0} & \olhsA{0} & \olhsC{0} \\
                        6 & \olhsC{1} & \olhsA{1} & \olhsC{1} & \olhsC{0} \\
                        7 & \olhsC{1} & \olhsB{2} & \olhsA{0} & \olhsB{1} \\
                        8 & \olhsC{1} & \olhsC{0} & \olhsB{2} & \olhsA{2} \\
                \end{tabular}
\end{center}

	\item From the construction above, in every column each symbol $k \in \{0,\dots,p-1\}$ appears exactly $N/q=p$ times, resulting in a set of points with $p$ samples in each subinterval $[k/p,(k+1)/p)$, which does not satisfy the desired LHS properties. One thus replaces the occurrence of each symbol with a random permutation on finer levels as follows. %
	Following \citet{Tang01121993}, we refine %
	each entry
	into $p$ subintervals of width $1/N$. Consider the $j$-th column and let $R_{j}^k$ be the set of rows with the symbol $k$, we randomly replace these entries with elements chosen at random from the set %
	$\{kp, kp+1,\dots,kp+p-1\}$. In our example, each column stores 3 sets of 3 symbols, %
	which are refined by independently assigning a random level from $\{0,\dots,8\}$. For instance for column $j=0$ the mapping is:
\begin{center}

                \begin{tabular}{cccc}
                        $k$ & Rows with $\tilde A_{i0}=k$ & Fine levels & Assigned \\
                        \midrule
                        \olhsB{0} & 3,\ 4,\ 5 & $\{0,1,2\}$ & $(1,2,0)$ \\
                        \olhsC{1} & 6,\ 7,\ 8 & $\{3,4,5\}$ & $(5,3,4)$ \\
                        \olhsA{2} & 0,\ 1,\ 2 & $\{6,7,8\}$ & $(6,8,7)$ \\
\end{tabular}
\end{center}

This refinement  results in the following table:
\begin{center}
                 \begin{tabular}{ccccc}
                        $i$ & $L_{i0}$ & $L_{i1}$ & $L_{i2}$ & $L_{i3}$ \\
                        \midrule
                        0 & \olhsA{6} & \olhsA{3} & \olhsA{1} & \olhsA{6} \\
                        1 & \olhsA{8} & \olhsB{7} & \olhsB{6} & \olhsC{2} \\
                        2 & \olhsA{7} & \olhsC{2} & \olhsC{5} & \olhsB{4} \\
                        3 & \olhsB{1} & \olhsA{4} & \olhsB{8} & \olhsB{5} \\
                        4 & \olhsB{2} & \olhsB{6} & \olhsC{4} & \olhsA{7} \\
                        5 & \olhsB{0} & \olhsC{1} & \olhsA{0} & \olhsC{0} \\
                        6 & \olhsC{5} & \olhsA{5} & \olhsC{3} & \olhsC{1} \\
                        7 & \olhsC{3} & \olhsB{8} & \olhsA{2} & \olhsB{3} \\
                        8 & \olhsC{4} & \olhsC{0} & \olhsB{7} & \olhsA{8} \\
                \end{tabular}
\end{center}

	\item Finally, add an independent uniform jitter to randomize the position inside each cell and rescale each dimension to $[0,1)$. For our $p=3$ example, the final result is the following. 
\begin{center}
                \begin{tabular}{ccccc}
                        $i$ & $u_{i0}$ & $u_{i1}$ & $u_{i2}$ & $u_{i3}$ \\
                        \midrule
                        0 & \olhsA{0.708} & \olhsA{0.423} & \olhsA{0.128} & \olhsA{0.736} \\
                        1 & \olhsA{0.894} & \olhsB{0.827} & \olhsB{0.770} & \olhsC{0.253} \\
                        2 & \olhsA{0.857} & \olhsC{0.243} & \olhsC{0.618} & \olhsB{0.538} \\
                        3 & \olhsB{0.164} & \olhsA{0.519} & \olhsB{0.899} & \olhsB{0.592} \\
                        4 & \olhsB{0.322} & \olhsB{0.692} & \olhsC{0.528} & \olhsA{0.834} \\
                        5 & \olhsB{0.016} & \olhsC{0.176} & \olhsA{0.047} & \olhsC{0.008} \\
                        6 & \olhsC{0.626} & \olhsA{0.591} & \olhsC{0.431} & \olhsC{0.162} \\
                        7 & \olhsC{0.362} & \olhsB{0.994} & \olhsA{0.290} & \olhsB{0.354} \\
                        8 & \olhsC{0.532} & \olhsC{0.012} & \olhsB{0.816} & \olhsA{0.969} \\
                \end{tabular}
\end{center}
Note how every $9\times9$ %
	grid built from any pair of columns
                holds exactly one point per cell.

\item
In compact notation, the final element $u_{ij}$ is determined as
\begin{equation}
	u_{ij}=\frac{kp+\pi_j^k(i)+V_{ij}}{N}, \quad i \in R_j^k, V_{ij} \sim \mathcal{U}[0,1),
\end{equation}
with all permutations $\pi_j^k(i)$ and jitters $V_{ij}$ mutually independent. %
The result is a set of columns that are marginally uniform in one dimension and for which any two-dimensional subset remains an orthogonal array of strength 2. By selecting $d$ columns, we obtain an orthogonal array Latin hypercube set OA-LHS$\left(N=p^2, d\right)$ for which the 2-dimensional uniform marginalizations are guaranteed.
\end{enumerate}

\begin{figure}[t]
	\centering
	\includegraphics[width=0.9\columnwidth]{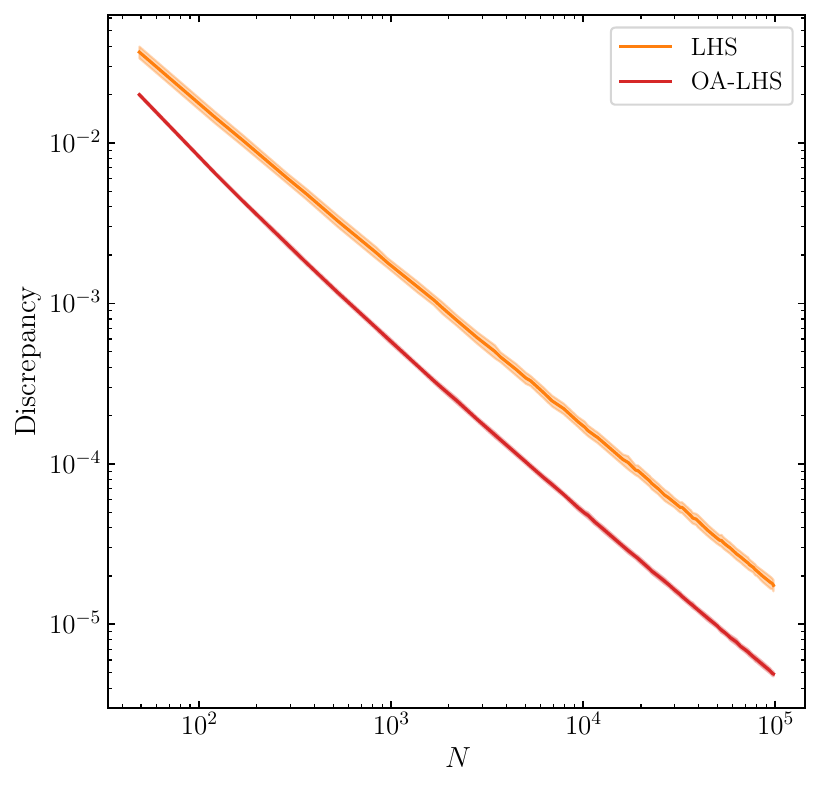}
	\caption{Average discrepancy of a sample set in $d=8$ dimensions as a function of the number of samples $N$. The orange (red) line shows how well the set generated using the LHS (OA-LHS) sampling technique fills the unit hypercube. Solid lines correspond to the average over 100 realizations, shaded areas are the standard deviation of the sets of samples.}
	\label{fig:two}
\end{figure}

An example is provided in the right panel of Fig.~\ref{fig:one}. Compared to standard LHS, exploiting orthogonal arrays allows for improved space-filling properties. To show this, we compute the centered discrepancy~\citep{ZHOU2013283} of a 8-dimensional sample set averaged over 100 realizations varying the number of draws. Results are shown in Fig.~\ref{fig:two}. For this particular metric, smaller values correspond to a more uniform coverage of the unit hypercube. The (minor) drawback in the construction described above is that it only applies to cases where the targeted number of samples $N$ is the square of a prime number. %

\subsection{Monte Carlo integrals}
Consider the following integral
\begin{equation}
	I=\int_{[0,1)^d}f(\boldsymbol{x})d\boldsymbol{x}\,\,.
\end{equation}
The plain Monte Carlo estimator of $I$  is
\begin{equation}
\hat{I}_{\rm MC}=\frac{1}{N}\sum_{i=1}^{N} f(\boldsymbol{u}_i)\,, \qquad \boldsymbol{u}_i\sim \mathcal{U}[0,1)^d\,,
\end{equation}
where $\mathcal{U}$ indicates the uniform distribution. This estimator is unbiased and its variance is ${\rm Var}[\hat{I}_{\rm MC}] = \sigma^2/N$ where $\sigma^2={\rm Var}\left[f(\boldsymbol{u})\right]$.

\citet{ef76b040-2f28-37ba-b0c4-02ed99573416} first showed that the variance of the estimator
\begin{equation}
\hat{I}_{\rm LHS}=\frac{1}{N}\sum_{i=1}^{N} f(\boldsymbol{u}_i)\,, \qquad \boldsymbol{u}_i\sim {\rm LHS}(N,d)\,,
\end{equation}
computed with LHS samples is not larger than the standard Monte Carlo approach if the integrand is monotone in each of its arguments. A generalization of this argument was provided by \citet{fbe9475b-1452-3229-a386-04540442a9c3}, who employed the functional ANOVA decomposition~\citep{55313c38-914c-3260-914e-410e7a9dbf9d,dca15e5b-b3f7-3417-8555-955fe36eb045,Sobol_1990aa}.
The key result is that, for finite values of $N$, it is mathematically proven  that the variance of an LHS-based Monte Carlo estimator is no larger than the variance of a plain Monte Carlo estimator with $N-1$ points, i.e.~\citep{885abe14-fb0a-3d11-96b5-11bed81c7313}.
\begin{equation}
{\rm Var}\left[\hat{I}_{\rm LHS}\right]\leq\frac{\sigma^2}{(N-1)} \simeq \frac{\sigma^2}{N} = {\rm Var}\left[\hat{I}_{\rm Mc}\right]\,.
\end{equation}
In short, using LHS rather than uniform random sampling guarantees an equal or lower variance.

In more detail, let us consider a square-integrable function in $d$ variables $f(\boldsymbol{x}) \colon[0,1)^d \mapsto \mathbb{R}$ and a set of $\boldsymbol{x}_j \in  [0,1)^d$ independent and uniformly distributed. Then, for $\alpha \subseteq \{0,\dots,d-1\}$ \citep{fbe9475b-1452-3229-a386-04540442a9c3} %
\begin{equation}
	\label{eq:anova}
	\begin{split}
		f(\boldsymbol{x}) & = \sum_{\alpha} f_{\alpha}(\boldsymbol{x}_{\alpha}) \\
		& = f_{\mathtt 0}+\sum_j f_j(x_j)+\sum_{j<k} f_{jk}(x_j,x_k)+\dots\,\, ,
	\end{split}
\end{equation}
where %
\begin{align}
	f_{\mathtt 0} & = \int f(\boldsymbol{x}) d\boldsymbol{x}\,,\\
	f_j(x_j) & = \int f(\boldsymbol{x}) d\boldsymbol{x}_{-j} - f_{\mathtt 0}\,, %
	\\
	f_{jk}(x_j,x_k) & =  \int f(\boldsymbol{x}) d\boldsymbol{x}_{-\{j,k\}} - f_j(x_j)-f_k(x_k)-f_{\mathtt 0}\,. %
\end{align}
Here $f_{\mathtt 0}$ is the average over all the coordinates,  $f_j(\boldsymbol{x}_j)$ is the average over all of them except $x_j$, representing how the mean of $f$ changes as $\boldsymbol{x}_j$ varies, and $f_{jk}$ contains pairwise interactions. Higher-order terms are defined similarly. By imposing that each term averages to zero over each of its own variables, one can demonstrate that the decomposition is unique; moreover, the variance becomes additive and can be written as %
\begin{equation}
	\label{eq:sigma_lhs}
	\sigma^2 =  \sum_j \sigma^2_j+\sum_{j<k} \sigma^2_{jk}+\dots\,\, ,
\end{equation}
where $\sigma^2_j={\rm Var}\left[f_j(x_j)\right]$ and $\sigma^2_{jk}={\rm Var}\left[f_{jk}(x_j,x_k)\right]$.
\citet{fbe9475b-1452-3229-a386-04540442a9c3} proved that the variance of a Monte Carlo estimator based on samples drawn following the LHS procedure has the following $N\to \infty$ limit %

\begin{equation}
\label{smallervar}
	{\rm Var}\Bigl[\hat{I}_{\rm LHS}\Bigr] \simeq \frac{1}{N}\biggl(\sigma^2- \sum_j \sigma^2_j\biggr)\,. %
\end{equation}
Therefore, LHS asymptotically removes the variance associated with one-dimensional contributions. The resulting variance reduction depends on how closely the integrand can be approximated by an additive function, that is, a constant plus a sum of functions that each depend on a single coordinate: $f(\boldsymbol{x})\approx {\rm const.}+\sum_j f_j(x_j)$. %
LHS offers no asymptotic advantage for a function whose variance lies mostly in interactions {between coordinates}.

 Using an OA-LHS {of strenght 2} offers additional gain as it can remove the two-factor interactions~\citep{Tang01121993} %

\begin{equation}
	\label{eq:sigma_oalhs}
	{\rm Var}\Bigl[\hat{I}_{\rm OA-LHS}\Bigr] \simeq \frac{1}{N}\biggl(\sigma^2 - \sum_j \sigma^2_j - \sum_{j<k} \sigma^2_{jk} \biggr)\,\, .
\end{equation}

\begin{figure*}[ht]
	\centering
	\includegraphics[width=0.96\columnwidth]{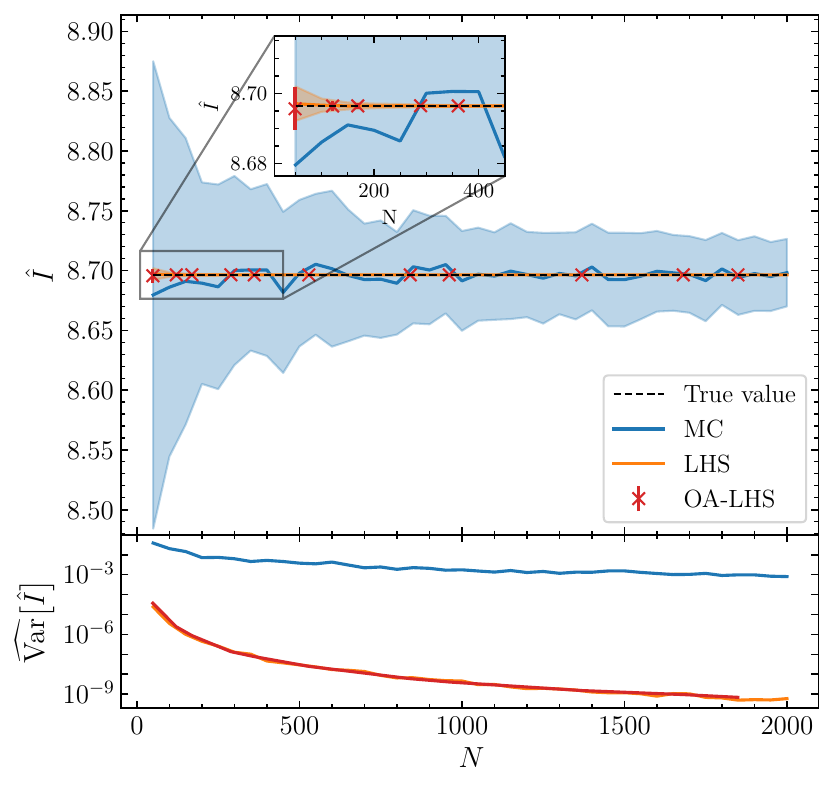}\hfill
	\includegraphics[width=0.96\columnwidth]{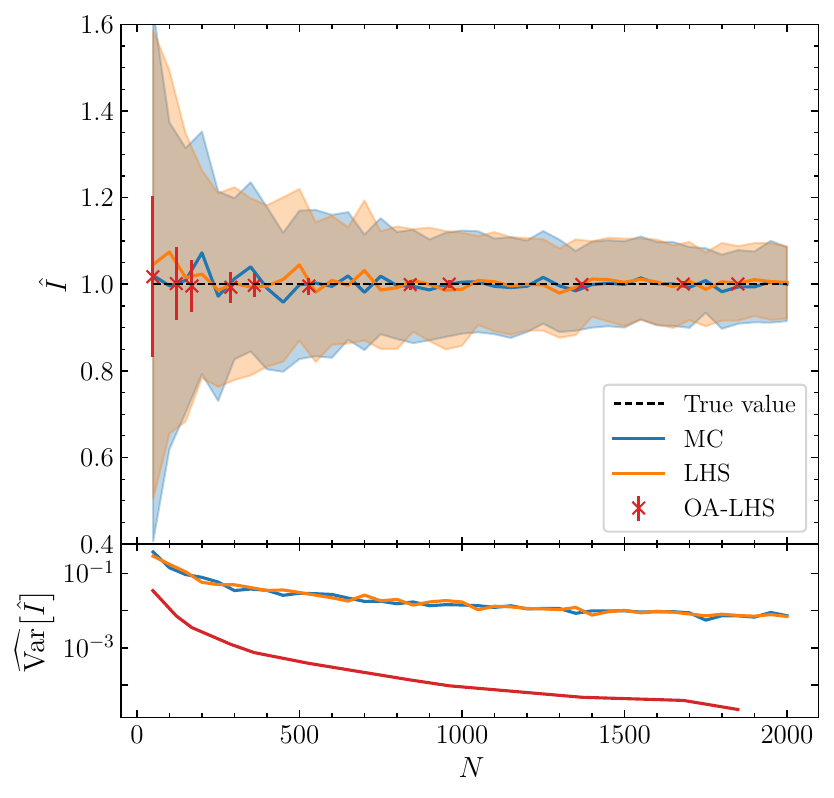}
	\caption{Left panel: Estimate and variance of the integral in Eq.~(\ref{app:eq_one}) with $\alpha_j=[0.5, 1, 2, 3, 5, 8, 12, 20]$. Right panel: Estimate and variance of the integral in Eq.~(\ref{eq:app_two}). The top panels show estimators computed via plain Monte Carlo (blue), LHS (orange) and OA-LHS (red) as a function of the number of drawn samples $N$. Sampling the latter is possible only when $N$ is equal to the square of a prime number. The true value is indicated by a dashed black line. Shaded areas and error bars represent the standard deviation computed on the 100 realizations. The inset in the top-left panel highlights the accuracy of the two Latin hypercube-based estimators even at low $N$. The bottom panels show the (estimated) variance of the three estimators ${\rm Var}\bigl[\hat{I}\bigr]$ as a function of $N$. %
	}
	\label{fig:app_one}
\end{figure*}

\section{Examples}
\label{examples}
We now present some examples to show the impact on the variance of Monte Carlo integrals by imposing one- and two-dimensional stratification using LHS and OA-LHS, respectively.

First, we consider the following integral on the unit hypercube in 8 dimensions
\begin{equation}
	\label{app:eq_one}
	I = \int_{[0,1)^8} \ln\prod\left(1+\boldsymbol{\alpha}\cdot\boldsymbol{x}\right) d\boldsymbol{x} =
\sum_{j=1}^{8}\int_0^1\ln(1+\alpha_j x_j)\,dx_j\,,
\end{equation}
with $\alpha_j > -1$. The integrand is a manifestly additive function. The true value of this integral can be computed analytically
\begin{equation}
		I_{\rm true} =
		\sum_{i=1}^{8}
		\left[\frac{1+\alpha_j}{\alpha_j}\ln(1+\alpha_j)-1\right]\,.
\end{equation}
We estimate the value of $I$ using the three sampling methods introduced earlier: standard Monte Carlo, LHS, and OA-LHS. We vary the number of samples drawn and compute the variance from 100 repeated realizations for each sample size; results are shown in the left panel of Fig.~\ref{fig:app_one}. The two Latin hypercube-based estimators greatly reduce the variance of the Monte Carlo integral due to the impact of the single-term variance in Eqs.~(\ref{smallervar}) and~(\ref{eq:sigma_oalhs}).

Our second example illustrates the effect of pairwise dependence in the functional ANOVA decomposition. Consider the integral
\begin{equation}
	\label{eq:app_two}
I = \int_{[0,1)^8} \bigg\{1+\sum_{j < k} \cos\left[2\pi(x_j-x_k)\right]\bigg\} \mathrm{d}\boldsymbol{x}\,,
\end{equation}
with $I_{\rm true}=1$. Results are shown in the right panel of Fig.~\ref{fig:app_one}. The one-dimensional stratification of LHS is not suited to capturing pairwise interactions, so the resulting variance is consistent with that obtained using standard Monte Carlo sampling. In contrast, the two-dimensional stratification of OA-LHS captures these interactions and reduces the overall variance. This reduction occurs because the integrand depends on interactions between pairs of coordinates; cf. Eq.~(\ref{eq:sigma_oalhs}).

\begin{table*}[t]
	\centering \small
	\setlength{\tabcolsep}{4pt}
	\renewcommand{\arraystretch}{1.15}
	\caption{Reference distribution $p_{\rm inj}(\theta)$ and its mapping onto
the coordinates $u_j$ of the unit hypercube. Masses are here defined in the source
frame, then converted to the detector frame %
when computing GW signals. %
}
	\label{tab:inj}
	\begin{tabular}{llll}
	\midrule
		\bf Coord. & \bf Symbol &\bf Parameter& \bf Distribution \\
		\midrule
		$u_0$ & $m_1$ &Primary mass & Power-law on $[1,1000]\,M_\odot$: slopes $0,-4,-1,-4,-1$, breaks at $3, 8, 50, 200\,M_\odot$ \\
		$u_1$ & $m_2\,|\,m_1$ & Secondary mass& $p(m_2\,|\,m_1) \propto m_2$ in $[1\,M_\odot, m_1]$ \\
		$u_2,\ u_5$ &  $\chi_{1},\ \chi_2$ &Spin magnitudes& Normal, zero mean, standard deviation $0.5$, truncated to $[0,1]$ \\
		$u_3,\ u_6$ & $\cos\theta_{1},\ \cos\theta_2$ &Spin polar angles& $70\%$ uniform on $[-1,1]$ $+$ $30\%$ Beta$(1,4)$ peaked at $\cos\theta = 1$ \\
		$u_4,\ u_7$ & $\phi_{1},\ \phi_2$ & Spin azimuthal angles &Uniform on $[0,2\pi)$ \\
		$u_8$ & $z$ &Redshift& $p(z) \propto \dd V_c/\dd z$ on $[0, z_{\max}]$, $z_{\max} = $ 2 \\
		$u_9$ & $\cos\theta_{JN}$ &Binary inclination& Uniform on $[-1,1]$ \\
		$u_{10}$ & $\phi_c$ & Phase at coalescence&Uniform on $[0,2\pi)$ \\
		$u_{11}$ & $\alpha$ & Right ascension &Uniform on $[0,2\pi)$ \\
		$u_{12}$ & $\sin\delta$ & Declination& Uniform on $[-1,1]$ \\
		$u_{13}$ & $\psi$ & Polarization & Uniform on $[0,\pi)$ \\
		$u_{14}$ & $t_c$ & Time at coalescence &Uniform within the observing window from 04/30/23 to 01/31/24 \\
		\bottomrule
	\end{tabular}
\end{table*}

\section{Gravitational-wave application}
\label{selectionfunction}

Here we apply the LHS and OA-LHS sampling strategies to the estimation of the GW selection function $\xi(\lambda)$ from Eqs.~(\ref{xiintegral}) and (\ref{xihat}). Our goal is to assess whether these strategies reduce the estimator variance relative to independent random sampling, while keeping both the sampling distribution $p_{\rm inj}$ and the number of draws fixed.

To this end, we generate repeated sample sets for each strategy and estimate the variance from the scatter of the resulting integral estimates. Realistic injection campaigns in GW astronomy assess $p({\rm det} | \theta)$ by adding simulated signals to detector data and running the search pipelines~\citep{2018CQGra..35n5009T,2025PhRvD.112j2001E}. Repeating such campaigns for multiple sampling strategies is computationally prohibitive, so we adopt a faster, semi-analytic model of detectability~\citep{1993PhRvD..47.2198F,2024CQGra..41l5002G}. Importantly, all strategies are evaluated using the same detection model, and our focus is on the relative variance of the estimators, not on the specific value of~$\xi(\lambda)$.

\subsection{Source detectability}

For a GW interferometer with one-sided noise power spectral density $S_n(f)$ and frequency-domain strain $\tilde h(f)=F_+\tilde h_+(f)+F_\times\tilde h_\times(f)$, the optimal signal-to-noise ratio (SNR) is given by~\citep{1987thyg.book..330T}
\begin{equation}
	\label{eq:snr_opt}
	\rho_{\rm opt}^2 = 4\int_{f_{\rm min}}^{f_{\rm max}} \frac{|\tilde h(f)|^2}{S_n(f)}\,\dd f\,,
\end{equation}
where the antenna patterns $F_{+,\times}$ depend on the sky position, the polarization angle, and, through the Earth's rotation, the coalescence time. For a network of $N_{\rm D}$ detectors, the individual optimal SNRs are combined in quadrature, $\rho_{\rm opt,net}^2=\sum_{D=1}^{N_{\rm D}}\rho_{{\rm opt},D}^2$. Assuming independent, stationary Gaussian noise, the observed SNR is normally distributed with mean $\rho_{{\rm opt},D}$ and unit variance \citep{2007gwte.book.....M}. The  network SNR $\rho_{\rm obs,net}$, therefore follows a non-central $\chi$ distribution~\citep{2024CQGra..41l5002G}. Defining a detection by $\rho_{\rm obs,net}>\rho_{\rm th}$, the detection probability is the survival function of this distribution:
\begin{equation}
	\label{eq:pdet}
	p({\rm det}\,|\,\theta)=Q_{N_{\rm D}/2}\!\left[\rho_{\rm opt,net}(\theta),\,\rho_{\rm th}\right],
\end{equation}
where $Q_\nu$ is the generalized Marcum-$Q$ function of order $\nu$. Neglecting noise fluctuations instead yields the step-function approximation $\Theta(\rho_{\rm opt,net}-\rho_{\rm th})$.

We consider a network of $N_{\rm D}=3$ GW interferometers: LIGO Hanford, LIGO Livingston~\citep{2015CQGra..32g4001L}, and Virgo~\citep{2015CQGra..32b4001A}. We adopt a detection threshold of $\rho_{\rm th}=10$. Under this model, a source with $\rho_{\rm opt,net}=\rho_{\rm th}$ is detected with probability $\simeq 0.540$, while the 50\% detection point lies at $\rho_{\rm opt,net}\simeq 9.899$. The detection probability rises from 10\% to 90\% over an interval $\Delta\rho_{\rm opt,net}\simeq 2.576$. %

We draw injections from the reference distribution $p_{\rm inj}(\theta)$ used during LIGO/Virgo's fourth observing run~\citep{2025PhRvD.112j2001E}, but reduce the upper redshift bound to $z_{\rm max}=2$. Each of the $d=15$ components of $\theta$ corresponds to one coordinate of the unit hypercube, and the physical parameters are obtained using inverse cumulative distribution functions, as summarized in Table~\ref{tab:inj}. Each coordinate is transformed independently, except for the secondary mass, which is drawn conditionally on the primary mass. The independently transformed parameters therefore inherit the stratification of the design in $[0,1)^{15}$. Luminosity distances are computed from redshift in a flat $\Lambda$CDM cosmology with $H_0=67.90\,{\rm km\,s^{-1}\,Mpc^{-1}}$, $\Omega_m=0.3065$, and $\Omega_\Lambda=1-\Omega_m$%
~\citep{2016A&A...594A..13P}.

We evaluate the selection function considering the \textsc{broken power law + 2 peaks} model for the masses, the \textsc{gaussian component spins} model for the BH spins, and the \textsc{power law redshift} model for $z$ from the GWTC-4 population analysis~\citep{2025arXiv250818083T}. We fix the hyperparameters $\lambda$ to their posterior medians. This population model specifies only a subset of parameters $\bar\theta\in \theta$, namely BH masses, spin magnitudes, spin directions, and redshift. The remaining parameters follow the same distribution in $p_{\rm pop}$ and $p_{\rm inj}$ and their densities cancel in the importance weights:
\begin{equation}
  \hat\xi(\lambda) = \frac{1}{N}\sum_{i=1}^{N}
  \frac{p_{\rm pop}(\bar\theta_i \,|\,\lambda)}{p_{\rm inj}(\bar \theta_i)}
  \, p({\rm det}\,|\,\theta_i)\,.
  \label{eq:xi_hat}
\end{equation}
The detection probability $p({\rm det} |\theta )$ still depends on all $15$ parameters. %
As a consistency check, we also evaluate the estimator with $p_{\rm pop} = p_{\rm inj}$, for which all weights are unity and $\hat\xi$ reduces to the mean detection probability of the injection set.

We generate GW signals using the precessing, higher-mode waveform model IMRPhenomXPHM~\citep{2021PhRvD.103j4056P}, as implemented in LALSimulation~\citep{lalsuite}. We adopt a frequency resolution of $\Delta f=1/8$~Hz between $f_{\rm min}=16$~Hz and $f_{\rm max}=2048$~Hz, with a reference frequency of $f_{\rm ref}=16$~Hz. We use the Advanced LIGO and Advanced Virgo design-sensitivity noise curves by~\cite{2020LRR....23....3A}. Antenna patterns are evaluated at each source's coalescence time, accounting consistently for the Earth's rotation throughout the observing window.

\subsection{Selection function}

We compare three sampling designs in $[0,1)^{15}$: standard Monte Carlo with independent uniform draws, LHS, and strength-2 OA-LHS, all mapped using the transformations in Table~\ref{tab:inj}. For each design, we vary the number of draws $N$, noting that our OA-LHS construction requires $N=p^2$, where $p\geq d-1$ is prime. For OA-LHS, we randomly select $15$ of the $p+1$ columns of the orthogonal array.
For each sampling strategy and sample size, we generate $R=200$ independent realizations, using independent permutations and jitter for LHS and OA-LHS.

\begin{figure}[t]
	\centering
	\includegraphics[width=1.0\columnwidth]{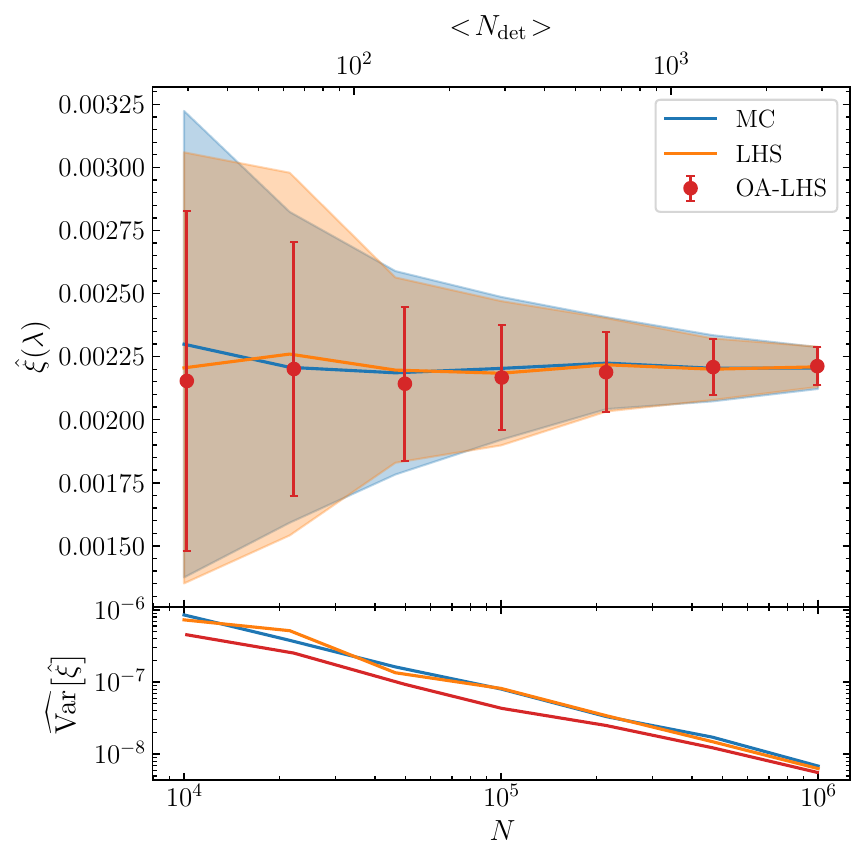}
	\caption{Selection function of GW astronomy estimated with three designs.
	The top panel shows estimators computed via plain Monte Carlo (blue), LHS (orange) and OA-LHS (red) as a function of the number of drawn samples $N$. %
	Sampling the latter is possible only when $N$ is equal to the square of a prime number.  The top axis shows the average number of detected binaries $\langle N_{\rm det} \rangle$. Shaded areas and error bars represent the standard deviation computed on the 200 realizations. The bottom panel shows the (estimated) variance of the three estimators $\widehat{\rm Var}\Bigl[\hat{\xi}\Bigr]$ as a function of $N$.
	}
	\label{fig:xi_results}

\end{figure}

Figure~\ref{fig:xi_results} shows the estimates of $\xi(\lambda)$ and their variances as functions of $N$. The three methods yield consistent estimates at every sample size, as expected for unbiased estimators. In all cases, their standard deviations decrease as $N^{-1/2}$. Although the OA-LHS estimator shows visible gains at the smaller values of $N$ considered here, the standard deviations of the three estimators become increasingly similar as $N$ increases.

To quantify these differences, we compare the variances of the estimators with a hypothesis test.
We first estimate the variance of each estimator from the scatter across the $R$ independent realizations,
\begin{equation}
	\label{eq:var_hat}
	\widehat{\rm Var}\big[\hat\xi\big] = \frac{1}{R-1}\sum_{r=1}^{R}\left(\hat\xi_r - \frac{1}{R}\sum_{r=1}^{R}\hat\xi_r\,\right)^2,
\end{equation}
{Assuming that the estimates $\hat\xi_r$ are approximately Gaussian,} the relative standard error of the estimated variance is $\sqrt{2/(R-1)}\simeq10\%$.
Under the null hypothesis that a LHS, OA-LHS design has the same variance as MC, the ratio $\mathcal{R}=\widehat{\rm Var}_{\rm LHS, OA-LHS}/\widehat{\rm Var}_{\rm MC}$, follows an $F$ distribution with $(R-1,R-1)$ degrees of freedom, whose central 95\% interval is $[0.76,1.32]$. A variance reduction is detected at the 5\% significance level only if
$\mathcal{R}<0.79$ (one-sided test): with this number of realizations, only reductions larger
than about 20\% can be resolved.%

For the estimates of $\xi(\lambda)$ shown in Fig.~\ref{fig:xi_results} we find $\mathcal{R}_{\rm LHS}=0.96$ and $\mathcal{R}_{\rm OA\text{-}LHS}=0.89$ %
at $N=994009$ injected binaries (which is the square of a prime number closest to $N=10^6$). %
With $R=200$ realizations, both ratios are consistent with unity at the 95\% confidence level. We therefore find no statistically significant difference between the variances of the three sampling methods for this application.

\subsection{Comments}

Although the gains we find are marginal, we stress that LHS-based estimators are {mathematically guaranteed} to have an asymptotic variance no greater than that of standard Monte Carlo. When computing the GW selection function, the computational cost is overwhelmingly dominated by the injections themselves. The additional cost of generating LHS and OA-LHS designs is therefore negligible, making even small variance reductions effectively free and worth considering.

All else being equal, including the number of samples $N$ and the injection distribution $p_{\rm inj}$, we recommend drawing the next set of LIGO/Virgo injections using OA-LHS (with the unusual requirement that the number of injections will need to be the square of a prime number). In addition, the injection distribution $p_{\rm inj}$ could be tailored to better exploit the variance reduction offered by LHS at a fixed number of injections $N$. 

This limited gain can be understood through a variance-based sensitivity analysis. We estimate the total contributions from single-parameter terms and pairwise interactions in Eqs.~(\ref{smallervar}) and (\ref{eq:sigma_oalhs}). For the integrand entering the selection function $\xi(\lambda)$, the normalized contributions are $\sum_i\sigma_i^2/\sigma^2=0.017$ and $\sum_{i<j}\sigma_{ij}^2/\sigma^2=0.124$, with the latter dominated by the interaction between primary mass and redshift. Thus, only 1.7\% of the variance is additive, while the remaining 85.9\% resides in interactions of third and higher order, which are not removed asymptotically by LHS or strength-2 OA-LHS.

Here we have applied LHS only to the selection function, rather than to the integrals over single-event measurement uncertainties that also enter GW population inference~\citep{2019MNRAS.486.1086M,2022hgwa.bookE..45V}. These take the form
\begin{equation}
\int
\frac{p(\theta | d_i)}{\pi(\theta)}
p_{\rm pop}(\theta | \lambda) \mathrm{d}\theta,
\end{equation}
where $p(\theta \mid d_i)$ is the posterior distribution for event $i$ obtained with reference prior $\pi(\theta)$. These integrals are typically evaluated using posterior samples $\theta \sim p(\theta \mid d_i)$ produced by a Bayesian parameter-estimation pipeline, so the sampling design is inherited from that analysis. An alternative, less explored approach evaluates these integrals using samples drawn from the population model, $\theta \sim p_{\rm pop}(\theta \mid \lambda)$, together with an approximation to the single-event likelihood \citep{2022ApJ...926...79G,2020ApJ...893...35D,2026arXiv261030187Y,desantiprep} . Space-filling designs such as LHS could play a role in this setting. Finally, sampling the full, non-marginalized population posterior avoids explicitly evaluating these single-event integrals \citep{2025PhRvD.111j3012M}, leaving only the selection-function integral considered in this paper.

We have evaluated the selection function $\xi(\lambda)$ for a single population model $p_{\rm pop}(\theta|\lambda)$ and at a single value of $\lambda$, corresponding to the marginal posterior medians from~\citet{2025arXiv250818083T}. Assessing the Monte Carlo variance across different population models and throughout their respective posterior distributions would provide a more comprehensive picture of the potential variance reduction achievable with LHS and OA-LHS.

\section{Conclusions}
\label{conclude}
When applied to Monte Carlo integration, space-filling designs can reduce the asymptotic variance of the resulting estimator. LHS enforces one-dimensional stratification, while orthogonal arrays extend this property to higher-dimensional marginals. In particular, strength-2 orthogonal arrays can be constructed algorithmically and used to stratify all two-dimensional projections. The resulting OA-LHS estimator can reduce variance by orders of magnitude when the integrand is dominated by additive contributions or pairwise interactions between coordinates.

A potential drawback of LHS designs is that the number of samples $N$ needs to be specified beforehand. Adding samples to an existing LHS design does not, in general, preserve the required stratification properties, though expansion strategies have been developed to minimize this effect \citep{Xiong01082009, CROMBECQ2011683, vovrechovsky2015hierarchical, SHEIKHOLESLAMI2017109, Li02092017, Zhou2019,2025SoftX..3102294B}.

Reducing Monte Carlo variance is particularly important in statistical inference, where noisy estimates of likelihood terms can hinder convergence. This is the case in GW population inference, where the accuracy of selection-effect estimates is a current bottleneck~\citep{2023MNRAS.526.3495T}. Although we find negligible gains with the current settings used in the field, these methods offer potential improvements ``for free,'' without requiring additional injections. More broadly, we hope that this practical introduction to LHS and OA-LHS will inspire further applications in physics and astronomy.

\section*{Acknowledgements}

We thank
Tristan Bruel,
Jack Heinzel, 
Matthew Mould, 
Alexandre Toubiana, 
and 
Salvatore Vitale
for discussions.
M.B. and D.G. are supported by
ERC Starting Grant No.~945155--GWmining,
Cariplo Foundation Grant No.~2021-0555,
Italian-French University (UIF/UFI) Grant No.~2025-C3-386,
MUR Grant ``Progetto Dipartimenti di Eccellenza 2023-2027'' (BiCoQ),
and the INFN TEONGRAV initiative.
D.G. is supported by MUR Young Researchers Grant No. SOE2024-0000125.
Computational work was performed
at CINECA with allocations through INFN and the University of Milano-Bicocca.

\bibliographystyle{elsarticle-authoryear-links}
\bibliography{lhspop}

\end{document}